\documentclass{article}

\pdfoutput=1
\usepackage[english]{babel}
\usepackage[margin=1in]{geometry}
\usepackage{multicol}
\usepackage{background}
\usepackage{longtable}
\usepackage{graphicx}

\SetBgScale{4}
\SetBgColor{gray}
\SetBgAngle{90}
\SetBgContents{arXiv: To be added}
\SetBgPosition{-1.5,-10}

\begin{document}

{\centering

{\bfseries\Large Resource Management in Cloud Computing: Classification and Taxonomy \bigskip}

Swapnil M Parikh\textsuperscript{1}, Dr. Narendra M Patel\textsuperscript{2}, Dr. Harshadkumar B Prajapati\textsuperscript{3} \\
{\itshape
\textsuperscript{1}PhD Scholar, Department of Computer Engineering, C U Shah University, Wadhwan and Assistant Professor, Department of Computer Science and Technology, BITS Edu Campus, Varnama, Vadodara, Gujarat, India. \\
\textsuperscript{2}Associate Professor, Department of Computer Engineering, Birla Vishvakarma Mahavidyalaya, Vallabh Vidyanagar, Gujarat, India. \\
\textsuperscript{3}Associate Professor, Department of Information Technology, Dharmsinh Desai University, Nadiad, Gujarat, India. \\
\normalfont (Dated: February 05, 2017)

}
}

\begin{abstract}
Cloud Computing is a new era of remote computing / Internet based computing where one can access their personal resources easily from any computer through Internet. Cloud delivers computing as a utility as it is available to the cloud consumers on demand. It is a simple pay-per-use consumer-provider service model. It contains large number of shared resources. So Resource Management is always a major issue in cloud computing like any other computing paradigm. Due to the availability of finite resources it is very challenging for cloud providers to provide all the requested resources. From the cloud providers perspective cloud resources must be allocated in a fair and efficient manner. Research Survey is not available from the perspective of resource management as a process in cloud computing. So this research paper provides a detailed sequential view / steps on resource management in cloud computing. Firstly this research paper classifies various resources in cloud computing. It also gives taxonomy on resource management in cloud computing through which one can do further research. Lastly comparisons on various resource management algorithms has been presented.

\bigskip


\end{abstract}

\begin{multicols}{2}
\section{Introduction}
Because of the advancement in Information and Communication Technology (ICT) over the past few years, Computing has been considered as a utility like water, electricity, gas and telephony. These utilities are available to the consumers based on their requirement at any time. Consumers pay for these services to the service providers based on their usage \cite{2Armbrust:2010:VCC:1721654.1721672,3buyya2009cloud,17manvi2014resource}.\\

Like all the other existing utilities, Computing utility is the basic computing service that meets the day to day needs of the general community. To deliver this vision, a number of computing paradigms have been proposed, of which the latest one is known as Cloud Computing. \\

Cloud is nothing but large pool of easily accessible and usable virtual resources. Cloud computing is a service provision model which provides various kinds of agile and effective services to the consumers where everything is considered as a service \cite{2Armbrust:2010:VCC:1721654.1721672,3buyya2009cloud,17manvi2014resource,9pallis2010cloud}. \\

Resource management is always a major issue at various computing areas. In cloud computing various cloud consumers demand variety of services as per their dynamically changing needs. So it is the job of cloud computing to avail all the demanded services to the cloud consumers. But due to the availability of finite resources it is very difficult for cloud providers to provide all the demanded services in time. From the cloud providers perspective cloud resources must be allocated in a fair manner. So, it is a vital issue to meet cloud consumers QoS requirements and satisfaction\cite{12parikh2013survey}. \\

Traditional resource management techniques are not adequate for cloud computing as they are based on virtualization technology with distributed nature. Cloud computing introduces new challenges for resource management due to heterogeneity in hardware capabilities, on-demand service model, pay per use model and guarantee to meet QoS \cite{26ben2012resource,25gonccalves2011resource,28xu2016building,27yuan2011efficient}.\\

Below mentioned are our major contributions for this research paper:
\begin{enumerate}
	\item 
	Firstly, this research paper classifies cloud resources based on utility.
	\item
	Secondly, this research paper gives a taxonomy on cloud resource management. The taxonomy is presented as a whole sequential process in two phases.
	\item
	Lastly this research paper presents comparisons on various resource management algorithms with their techniques, type of algorithm and research issues.
\end{enumerate} 

The rest of the paper is organized as follows: Section 2 discusses fundamentals of cloud computing. Section 3 presents classification on cloud resources. Section 4 gives taxonomy on cloud resource management. Section 5 shows comparisons on various resource management algorithms. Section 6 concludes this research work.

\section{Background}
Cloud is like a big black box, nothing inside the cloud is visible to the cloud consumers. Cloud delivers computing as a utility as it is available to the cloud consumers on demand. Cloud Computing is a simple pay-per-use consumer-provider service model.  \cite{2Armbrust:2010:VCC:1721654.1721672,3buyya2009cloud,17manvi2014resource,9pallis2010cloud}. \\

Below are the widely quoted definitions of Cloud Computing:
\begin{itemize}
	\item
	NIST\cite{1mell2011nist}: Cloud computing is a model for enabling ubiquitous, convenient, on-demand network access to a shared pool of configurable computing resources (e.g., networks, servers, storage, applications, and services) that can be rapidly provisioned and released with minimal management effort or service provider interaction.
	
	\item
	Ian Foster\cite{11foster2008cloud}: A large-scale distributed computing para-digm that is driven by economies of scale, in which a pool of abstracted, virtualized, dynamically-scalable, managed computing power, storage, platforms, and services are delivered on demand to external customers over the Internet.
	
	\item
	Rajkumar Buyya\cite{3buyya2009cloud}: A Cloud is a type of parallel and distributed system consisting of a collection of inter-connected and virtualized computers that are dynamically provisioned and presented as one or more unified computing resource(s) based on service-level agreements established through negotiation between the service provider and consumers.
	
\end{itemize}

Cloud computing is composed of three kinds of service models. These service models are based on the level and depth of the services provided by cloud computing \cite{11foster2008cloud,17manvi2014resource,1mell2011nist,10sadashiv2011cluster,5zhang2012research,6zissis2012addressing}.

\begin{enumerate}
	\item 
	Cloud Software as a Service (SaaS): In this service model, instead of using locally run applications the cloud consumer uses the cloud providers software services running on a cloud infrastructure. It is the job of cloud provider to maintain and manage the software services that are used by the cloud consumer. The cloud provider may charge according to software quantity and time usage. Salesforge.com and Customer Relationship Management (CRM) are the examples of such service model \cite{1mell2011nist,12parikh2013survey,4reddy2011research,7zhang2010hot,5zhang2012research,6zissis2012addressing}.
	
	\item
	Cloud Platform as a Service (PaaS): In this service model, the cloud platform offers an environment on which developers create and deploy applications. It provides platform where applications and services can run. The consumers do not need to take care of underlying cloud infrastructure including network, servers, operating system or storage but has a control over deployed application. Google Application Engine, Microsoft Azure and RightScale are the example of such model \cite{1mell2011nist,12parikh2013survey,4reddy2011research,5zhang2012research,6zissis2012addressing}.
	
	\item
	Cloud Infrastructure as a Service (IaaS): In this service model, cloud providers manage large set of computing resources such as storing and processing capability. Cloud consumer can control operating system; storage, deployed applications, and possibly limited control of select networking components (e.g., host firewalls). Sometimes it is also called as a Hardware as a Service (HaaS). The cost of the Hardware can be greatly reduced here. Amazon Web Services, Open Stack, Eucalyptus, GoGrid and Flexiscale offers IaaS \cite{1mell2011nist,12parikh2013survey,4reddy2011research,5zhang2012research,6zissis2012addressing}.
	
\end{enumerate}

In cloud computing various deployment models have been adopted based on their variation in physical location and distribution. Regardless of the services, clouds can be classified among four models as mentioned below.

\begin{enumerate}
	\item 
	Private Cloud: It is private to the organization. All the cloud services are managed by the organization people themselves or any third party vendors. In private cloud services are not provided to the general public. Private cloud may exist on premise or off premise \cite{8grossman2009case,1mell2011nist,5zhang2012research,6zissis2012addressing}. 
	
	\item
	Public or Hosted Cloud: All the cloud services managed by the organization are made available as in pay as you go manner to the general public. The business people can adopt such cloud to save their hardware and/or software cost. Public cloud may raise number of issues like data security, data management, performance, level of control etc \cite{8grossman2009case,1mell2011nist,5zhang2012research,6zissis2012addressing}.  
	
	\item
	Community Cloud: Here cloud is available to specific group of people or community. All the cloud services are shared by all these community people. Community cloud may exist on premise or off premise \cite{1mell2011nist,5zhang2012research,6zissis2012addressing}.  
	
	\item
	Hybrid Cloud: It is a combination of two or more cloud models mentioned above \cite{1mell2011nist,5zhang2012research,6zissis2012addressing}. 
\end{enumerate}

\includegraphics[width=\linewidth]{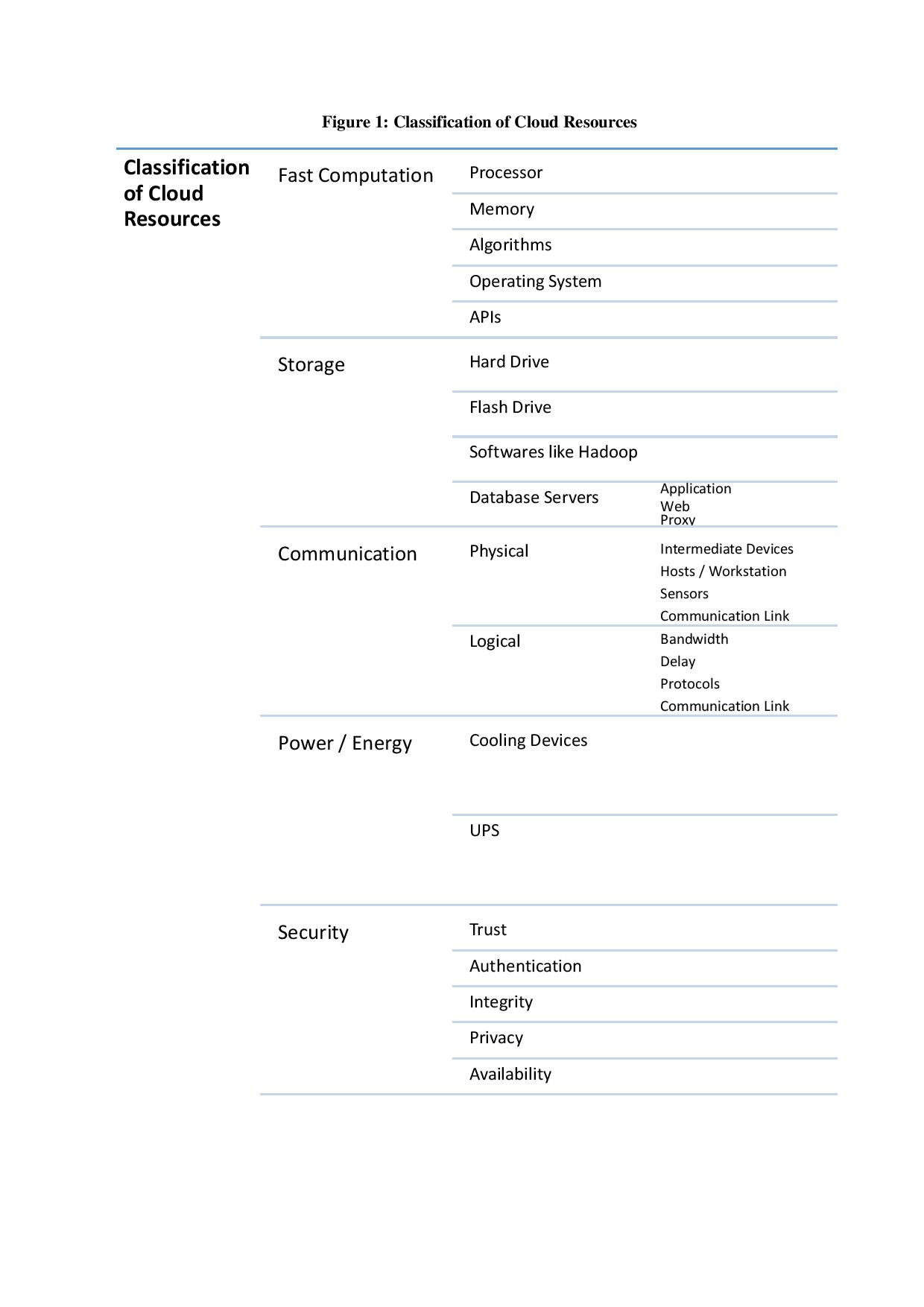}

\section{Classification of Cloud Resources}
Cloud computing provides a platform where resources are rented as a service to its cloud users / cloud consumers through Internet. So we can say that Cloud delivers computing as a utility as it is available to the cloud consumers on demand. \\

In cloud computing,  a resource can be any service which can be consumed by cloud users / cloud consumers. Different researchers have classified resource as physical resources and logical resources or hardware resources and software resources \cite{25gonccalves2011resource,16jennings2015resource,17manvi2014resource,15mustafa2015resource,19singh2015qos,18zhan2015cloud}. \\

In cloud computing, cloud providers manages various resources. As cloud computing is a utility based computing, this research paper classifies cloud resources based on utility. Figure 1 gives a detailed view on classification of resources in cloud computing.

\begin{enumerate}
	\item 
	Fast Computation Utility: This type of resources provide fast computational utility in cloud computing environment. Through fast computation utility cloud computing provides Computation as a Service (CaaS). Fast computation utility includes processing capability, size of memory, efficient algorithms etc \cite{16jennings2015resource,17manvi2014resource}.
	\item 
	Storage Utility: Instead of storing data at local storage device, we store them at storage device which is located at remote place. Storage utility consists of thousands of hard drives, flash drives, database servers etc. As computer systems are bound to fail over the period of time data redundancy is required here. Due to cloud's time variant service model storage utility needs to provide features like cloud elasticity \cite{16jennings2015resource,17manvi2014resource}. Through storage utility cloud computing provides Storage as a Service (StaaS).
	\item 
	Communication Utility: It can also be called as Network Utility or Network as a Service (NaaS). Fast computation utility and storage utility can not be thought without communication utility. Communication utility consists of physical (intermediate devices, hosts, sensors, physical communication link) and logical (bandwidth, delay, protocols, virtual communication link) resources. In cloud computing each and every service is provided through high speed Internet. So bandwidth and delay are most important from network point of view \cite{16jennings2015resource,17manvi2014resource}.
	\item 
	Power / Energy Utility: Now a days researchers are doing a lot of research work on energy efficient techniques in cloud computing. Energy cost can greatly be reduced by using power aware techniques. Due to thousands of data servers power consumption is very high in cloud computing. Cooling devices and UPS are at the center of these type of resources. They can also be considered as secondary resources \cite{21hameed2014survey,16jennings2015resource,17manvi2014resource}. 
	\item 
	Security Utility: Security is always a major issue in any computing area. Being cloud user we want highly reliable, trust-able, safe and secure cloud services \cite{17manvi2014resource}. 				
\end{enumerate}

\section{Taxonomy on Cloud Resource Management}

\includegraphics[width=\linewidth,height=2in]{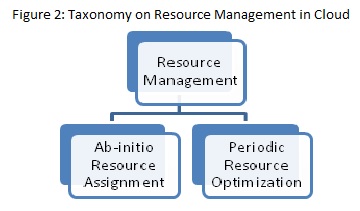}

The goal of resource management in cloud computing is to provide high availability of resources, sharing of resources, fulfilling time variant service model, providing efficiency and reliability on resource usage \cite{26ben2012resource,23mohamaddiah2014survey,28xu2016building,27yuan2011efficient}. \\

From the cloud computing perspective, resource management is a process which effectively and efficiently manages above mentioned resources as well as providing QoS guarantees to cloud consumers. This section gives Taxonomy on Cloud Resource Management (Refer Figure 2). The taxonomy is presented as a whole sequential process in two phases. 

\subsection{Phase 1: Ab-initio Resource Assignment}
It is initial resource assignment, in a manner that resources are requested by application (on behalf of cloud consumers) first time. Figure 3 shows several sequential steps which needs to be followed for completion of this phase \cite{16jennings2015resource,17manvi2014resource,15mustafa2015resource,19singh2015qos,20singh2016survey,18zhan2015cloud}.

\includegraphics[width=\linewidth,height=3.5in]{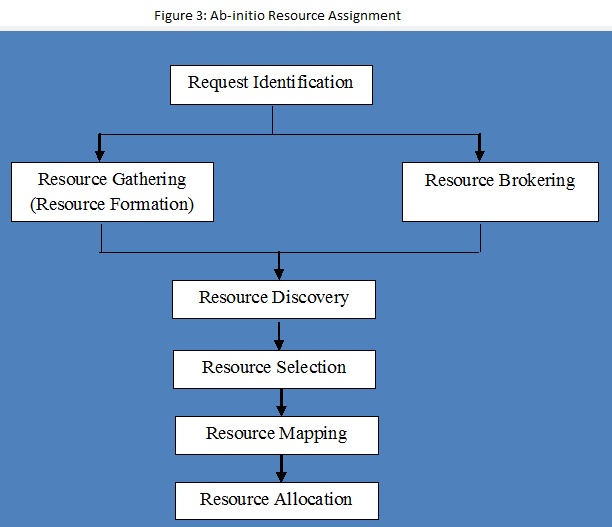}
	
\begin{enumerate}
	\item 
	Request Identification: This is the first and foremost step in Ab-initio Resource Assignment. In this step, various resources will be identified by cloud providers. 
	\item 
	Resource Gathering / Resource Formation: After identification of resources in step 1, gathering or formation of resources will take place. This step will identify available resources. This step may also prepare custom resources.
	\item 
	Resource Brokering: This step is negotiation of resources with cloud consumers to make sure that they are available as per requirement.
	\item 
	Resource Discovery: This step will logically group various resources as per the requirements of cloud consumers. 
	\item 
	Resource Selection: This step is to choose best resources among available resources for requirements provided by cloud consumers.
	\item 
	Resource Mapping: This step will map virtual resources with physical resources (like node, link etc) provided by cloud providers.
	\item 
	Resource Allocation: This step will allocate / distribute resources to the cloud consumers. It's main goal is to satisfy cloud consumers' need and revenue generation for cloud providers.
\end{enumerate}

\subsection{Phase 2: Periodic Resource Optimization}

\includegraphics[width=\linewidth,height=2in]{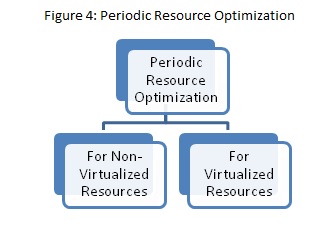}

As name suggest this is a phase where resource management is done at regular intervals once phase 1 is completed. Here periodic resource optimization is presented as a process for two different categories of resources which are non-virtualized resources and virtualized resources (Refer Figure 4)
\cite{22ergu2013analytic,16jennings2015resource,17manvi2014resource,15mustafa2015resource,19singh2015qos,20singh2016survey,18zhan2015cloud}. The non-virtualized resources are also called as physical resources. For both categories of resources, periodic resource optimization contains similar steps. The only difference is that virtualized resources can be assembled together as per the resource requirement and can be disassembled also. So periodic resource optimization for virtualized resources contains two steps more compared to non-virtualized resources which are Resource Bundling and Resource Fragmentation.

\begin{enumerate}
	\item 
	For Non-virtualized Resources (Refer Figure 5)
	
	\begin{enumerate}
		\item 
		Resource Monitoring: Resource Monitoring is the first and crucial step in Periodic Resource Optimization. Various non-virtualized cloud resources are monitored to analyze utilization of resources. This step will also monitor availability of free resources for future purpose. The major issue with cloud resource monitoring is to identify and define metrics/parameters for it. 
		\item
		Resource Modeling / Resource Prediction: This step will predict the various non-virtualized resources required by cloud consumers applications. This is one of the complex step as cloud resources are not uniform in nature. Due to this non uniformity, it is very difficult to predict resource requirement for peak periods and as well as for non-peak periods.
		\item
		Resource Brokering: This step is negotiation of non-virtualized resources with cloud consumers to make sure that they are available as per requirement.
		\item
		Resource Adaptation: As per the requirements of cloud consumers, non-virtualized cloud resources can be scaled up or scaled down. This step may increase cost from cloud providers perspective.
		\item
		Resource Reallocation: This step will reallocate / redistribute resources to the cloud consumers. It's main goal is to satisfy cloud consumers' need and revenue generation for cloud providers.
		\item
		Resource Pricing: It is one of the most important step from cloud providers and cloud consumers perspective. Based on cloud resource usage pricing will be done.
	\end{enumerate}

	\includegraphics[width=\linewidth,height=4in]{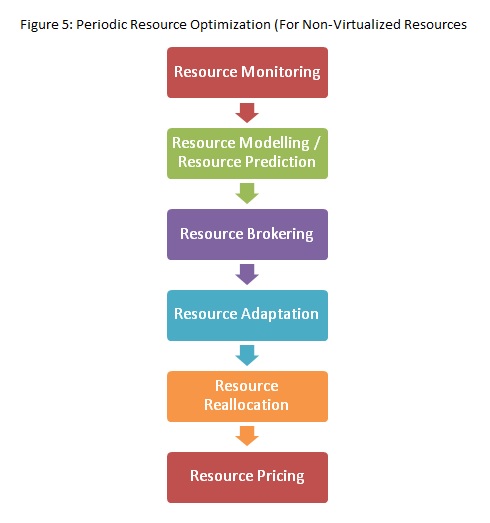}

	\item 
	For Virtualized Resources (Refer Figure 6)
	
	\includegraphics[width=\linewidth,height=4in]{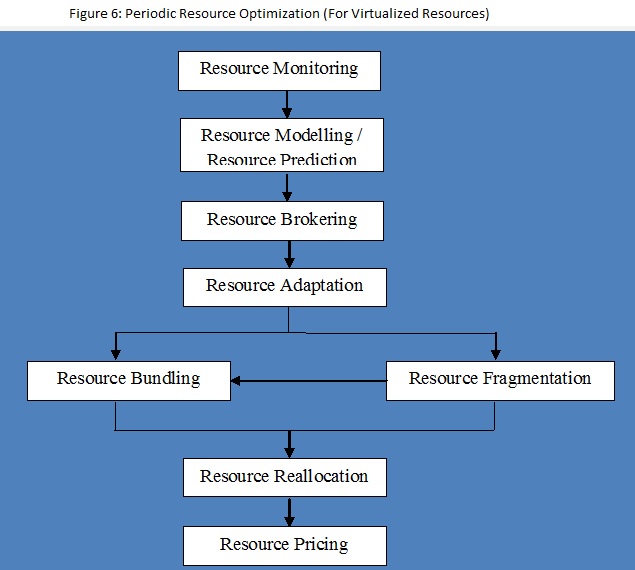}
		
	\begin{enumerate}
		\item 
		Resource Monitoring: Resource Monitoring is the first and crucial step in Periodic Resource Optimization. Various virtualized cloud resources are monitored to analyze utilization of resources. This step will also monitor availability of free resources for future purpose. The major issue with cloud resource monitoring is to identify and define metrics / parameters for it. 
		\item 
		Resource Modeling / Resource Prediction: This step will predict the various virtualized resources required by cloud consumers applications. This is one of the complex step as resources are not uniform in nature. Due to this non uniformity, it is very difficult to predict resource requirement for peak periods and as well as for non-peak periods.
		\item 
		Resource Brokering: This step is negotiation of virtualized resources with cloud consumers to make sure that they are available as per requirement.
		\item 
		Resource Adaptation: As per the requirements of cloud consumers, virtualized cloud resources can be scaled up or scaled down. This step may increase cost from cloud providers perspective.
		\item 
		Resource Bundling: As per the requirement various non-virtualized resources can be bundled into virtualized resources. 
		\item 
		Resource Fragmentation: Various virtualized resources needs to be fragmented to make non virtualized resources free. After this step various non-virtualized resources can be bundled in to virtualized resources as a part of resource bundling.
		\item 
		Resource Reallocation: This step will reallocate / redistribute resources to the cloud consumers. It's main goal is to satisfy cloud consumers' need and revenue generation for cloud providers.
		\item 
		Resource Pricing: It is one of the most important step from cloud providers and cloud consumers perspective. Based on cloud resource usage pricing will be done.
	\end{enumerate}
\end{enumerate}

\includegraphics[width=\linewidth,height=2in]{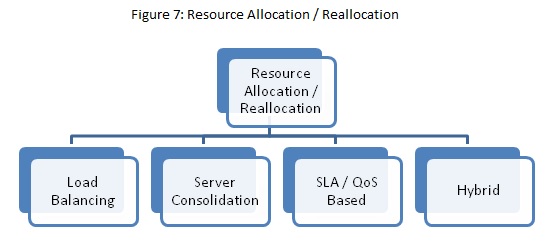}

Resource Allocation and Resource Reallocation can be done based on below mentioned broadly classified policies as shown in Figure 7:
\begin{enumerate}
	\item 
	Load Balancing
	\item
	Server Consolidation
	\item
	SLA / QoS based
	\item
	Hybrid
\end{enumerate}

\begin{table*}
	\centering
	{Table 1: Comparisons of Resource Management Algorithms
		\label{Table}} {
		
		\begin{tabular}{| c | p{2.2cm} | p{2.5cm} | p{2cm} | p{2cm} | p{2.5cm} |}
			\hline
			\textbf{Sr. No} & \textbf{Publication (Year)} & \textbf{Techniques / Algorithms} & \textbf{Tools and/or workload used} & \textbf{Type} & \textbf{Future work and/or gaps in existing technologies} \\
			\hline
			1 & Proceedings of the World Congress on Engineering and Computer Science (2011) \cite{29galloway2011power} & Power Aware Load Balancing Algorithm & Eucalyptus & Resource Allocation / Reallocation (Load Balancing) & Energy Savings not considered. \\
			\hline
			2 & 4th International IEEE Conference on Utility and Cloud Computing (2011) \cite{30zhang2011dynamic} & Dynamic Resource Allocation for Spot Instances & Amazon EC2 & Resource Allocation / Reallocation & Customer’s perspective and bidding behavior is not considered. \\
			\hline 
			3 & IEEE Transactions (2012) \cite{31papagianni2013optimal} & Optimal Allocation of Virtual Resources using Mixed Integer Programming (MIP) & Simulator for Controlling Virtual Infrastructures (CVI-Sim) & Resource Allocation / Reallocation & Implementation of proposed framework.
			\\
			\hline
			
			4 & ELSEVIER-Information Sciences (2014) \cite{32samimi2014combinatorial} & Combinatorial Double Auction Resource Allocation (CDARA) & CloudSim & Resource Allocation / Reallocation & Experiments were done on simulators, not on real environments. \\
			\hline
			5 & ELSEVIER-Procedia Computer Science (2016) \cite{33jha2016power} & Power and Load Aware VM Allocation Policy & CloudSim & Resource Allocation / Reallocation & Experiments were done on simulators, not on real environments. \\
			\hline
			6 & Springer (2010) \cite{34ge2010research} & Resource Monitoring Model for Cloud Computing & Linux C/C++ and Java & Resource Monitoring & Reliable Resource Discovery is future work. \\
			\hline
			7 & J Grid Computing – Springer (2015) \cite{35gutierrez2016iaasmon} & IaaSMon & Nagios / OpenStack & Resource Monitoring & Integration of both tools. \\
			\hline
			8 & IEEE/ACM (2010) \cite{36mihailescu2010dynamic} & Dynamic Resource Pricing & PlanetLab & Resource Pricing & Scalability is a issue. \\
			\hline
			
		\end{tabular}}
	\end{table*}
	
	\section{Summary and Comparisons of Resource Management Algorithms}
	Cloud resource management process is very complex in nature. In above sections, the whole cloud resource management process had been clearly divided among various steps / techniques which distinguishes all of them from one another. Below is the summary on various resource management techniques.
	
	\cite{29galloway2011power} presented Power Aware Load Balancing Algorithm (PALB) for IaaS cloud. Authors had designed algorithm in three segments. 1) Balancing Segment 2) Upscale Segment and 3) Downscale Segment. PALB maintains the status of all compute nodes and based on their usage, they decide the number of functional compute nodes.
	
	\cite{30zhang2011dynamic} proposed market driven resource allocation technique. Authors developed discrete event based VM scheduler for resource management. Authors used single provider scenario for spot instance service provided by Amazon EC2. After performing evaluation authors claim that average request waiting time is reduced.
	
	\cite{31papagianni2013optimal} had proposed a method for the efficient mapping of resource requests onto a shared substrate interconnecting various islands of computing resources, and adopt a heuristic methodology to address the problem.
	
	\cite{32samimi2014combinatorial} had proposed Combinatorial Double Auction Resource Allocation (CDARA) which is a market driven model for resource management in cloud computing. CDARA consists of seven communication phases. 1) Advertising and resource discovery 2) Generate bundles 3) Informing the end of auction 4) Winner determination 5) Resource allocation 6) Pricing model 7) Allocation of task and payment. Authors used CloudSim for simulation in cloud. It is an auction based model. 
	
	\cite{33jha2016power} proposed power and load aware resource allocation policy for hybrid cloud. Authors tried to minimize power consumption and maximize utilization of resources. Authors have developed two separate algorithms: 1) resource initialization and 2) resource allocation. Authors tested their algorithms with DVFS based scheduling technique.
	
	\cite{34ge2010research} had proposed resource monitoring model for virtual machine in cloud computing. Authors had monitored live working nodes static and dynamic information for future resource discovery and resource allocation models. Implementation was done using C/C++ and Java language.
	
	\cite{35gutierrez2016iaasmon} had proposed monitoring architecture for cloud computing. To achieve this, authors had done integration between resource monitoring tool and its resource discovery protocol. Implementation of the same is done in Java.
	
	\cite{36mihailescu2010dynamic} had majorly focused on dynamic resource pricing in cloud computing. Authors claim that a dynamic pricing policy is always able to balance the number of successful requests and the number of allocated resources depending on the market condition. So it achieves better economy efficiency. 
	
	Table 1 shows comparisons on above summarized resource management algorithms with their techniques, type of algorithm and research issues.

\section{Conclusions}
	Cloud computing enables cloud resources to be used as a utility. Through analyzing cloud computing for resource management, this research paper first focused on classifying cloud resources. After that taxonomy on cloud resource management was presented so that various research issues related to resource management can be identified based on various phases and stages mentioned in this paper. Lastly various research papers were reviewed for identifying research issues in cloud resource management. \\ 
	
	In summary, this research paper presents resource management in cloud computing as a sequential process of various techniques with their research issues. This research paper also concludes that efficient cloud resource management should meet criteria's like efficient utilization of resources, cost reduction from cloud providers perspective, energy / power reduction. \\
	
	There can be several future directions for this research work. One of the future work is to identify various techniques of resource allocation / reallocation through multi-objective optimization techniques. Moreover, novel optimized techniques have to be formulated that should accommodate above mentioned criteria.

\bibliographystyle{abbrv}
\bibliography{references}

\end{multicols}

\end{document}